\documentclass[aps,twocolumn,nofootinbib,showpacs]{revtex4}
\usepackage{bm,amssymb,graphicx}
\newcommand{\iR}{{\rm i}}

\newcommand{\Dket}[1]{\left|#1\right\rangle}
\newcommand{\Dbraket}[2]{\langle#1|#2\rangle}
\newcommand{\Qcommu}[2]{[#1,#2]}
\newcommand{\Qantico}[2]{\{#1,#2\}}

\begin{document}
\bibliographystyle{apsrev}

\title{BRST quantization of Yang-Mills theory: A purely Hamiltonian approach on Fock space}

\author{Hans Christian \"Ottinger}
\email[]{hco@mat.ethz.ch}
\homepage[]{www.polyphys.mat.ethz.ch}
\affiliation{ETH Z\"urich, Department of Materials, Polymer Physics, HCP F 47.2,
CH-8093 Z\"urich, Switzerland}

\date{\today}

\begin{abstract}
We develop the basic ideas and equations for the BRST quantization of Yang-Mills theories in an explicit Hamiltonian approach, without any reference to the Lagrangian approach at any stage of the development.
We present a new representation of ghost fields that combines desirable self-adjointness properties with canonical anticommutation relations for ghost creation and annihilation operators, thus enabling us to characterize the physical states on a well-defined Fock space. The Hamiltonian is constructed by piecing together simple BRST invariant operators to obtain a minimal invariant extension of the free theory. It is verified that the evolution equations implied by the resulting minimal Hamiltonian provide a quantum version of the classical Yang-Mills equations. The modifications and requirements for the inclusion of matter are discussed in detail.
\end{abstract}


% 03.70.+k   Theory of quantized fields (see also 11.10.-z Field theory)
% 12.20.-m   Quantum electrodynamics

\maketitle

\section{Introduction}
BRST quantization is a pivotal tool in developing theories of the fundamental interactions, where the acronym BRST refers to Becchi, Rouet, Stora \cite{BecchiRouetStora76} and Tyutin \cite{Tyutin75}. This method for handling constraints in the quantization of field theories usually requires a broad viewpoint because it covers a number of important aspects. The constraints are related to gauge symmetry, which suggests that a Lagrangian approach is preferable, in particular, as also Lorentz symmetry needs to be incorporated. By Noether's theorem, symmetries come with conserved quantities, which suggest to focus on time evolution and hence to favor the Hamiltonian approach. Practical calculations, for example in perturbation theory, are most conveniently done in terms of the path-integral formulation and hence on the Lagrangian side. The identification of physical states, which requires gauge fixing as an additional aspect of introducing constraints, is most naturally done on Hilbert space (as the space of all states). The issue of signed inner products, to be considered simultaneously with canonical inner products, requires particular attention in constructing the physical states and should clearly benefit from a simple and intuitive approach to BRST quantization.

The purpose of this paper is to show in the context of Yang-Mills theories how all the above facets can be handled entirely within the Hamiltonian approach, where explicit constructions on a suitable Fock space allow for a maximum of intuition. The focus on Fock space implies a (quantum) particle interpretation rather than a field idealization. The signed and canonical inner products are particularly transparent on Fock space. By including temporal and longitudinal in addition to transverse gauge bosons (we typically think of photons or gluons), Lorentz symmetry is enabled at the early stage of constructing the underlying Fock space. All symmetry arguments are based on the BRST charge, the construction of which relies on its role as the generator of BRST transformations, the quantum version of gauge transformations. Of course, the BRST charge has to respect also Lorentz symmetry.

In this paper, we present a new way of introducing ghost particles. Whereas one usually has to make the choice between natural anticommutation relations for the ghost creation and annihilation operators on the one hand (see, e.g., \cite{Nemeschanskyetal86}) and self-adjointness of the BRST charge on the other hand (see, e.g., \cite{KugoOjima78a}), we here propose a representation of ghost particles that combines both properties. This is a crucial advantage because ``In the non-Abelian case, the removal of unphysical gauge boson polarizations is more subtle [than in the Abelian case], and we have seen that it involves the ghosts in an essential way'' (see p.\,520 of \cite{PeskinSchroeder}). We simultaneously have a well-defined Fock space and the powerful tools required to select the physical states of a gauge theory in the BRST approach.

After constructing a number of simple BRST invariant operators, these operators can be used to build up the BRST invariant Hamiltonian. By piecing together invariant operators to reproduce the proper Hamiltonian of the free theory for vanishing interaction strength, one obtains the Hamiltonian of Yang-Mills theory as a minimal BRST invariant extension of the free theory. The validity of the Hamiltonian can be verified by comparing the time evolution implied by this Hamiltonian to the classical evolution equations. Matter can be included into the Hamiltonian approach with the help of the current algebra.

\newpage

\section{Classical Yang-Mills theory}\label{secclassYM}
Yang-Mills theory introduces antisymmetric fields $F^a_{\mu\nu}$ that are defined in terms of four-vector potentials $A^a_\mu$,
\begin{equation}\label{classYMfields}
   F^a_{\mu\nu} = \partial_\mu A^a_\nu - \partial_\nu A^a_\mu
   - g f^{abc} A^b_\mu A^c_\nu ,
\end{equation}
where the superscripts $a,b,c$ label the generators of the underlying symmetry group and the indices $\mu,\nu$ label the space-time components; the parameter $g$ is the strength of the interaction and the set of numbers $f^{abc}$ are the structure constants of the underlying Lie group. The field equations are given by
\begin{equation}\label{classYMevol}
   \partial^\mu F^a_{\mu\nu} - g f^{abc} A^{b \mu} F^c_{\mu\nu} = - J^a_\nu ,
\end{equation}
where the four-vector $J^a_\nu$ is an external current. The parameter $g$ usually occurs with the opposite sign because we here choose the signature $(-,+,+,+)$ for the Minkowski metric, contrary to the more common convention $(+,-,-,-)$; the four-vectors $\partial_\mu$ and $A^{a \mu}$ are independent of the signature of the Minkowski metric. The term $\tilde{J}^a_\nu = -g f^{abc} A^{b \mu} F^c_{\mu\nu}$ may be interpreted as the current carried by the gauge bosons. The structure constants can be assumed to be completely antisymmetric in the indices $a,b,c$ (see Sec.~15.4 of \cite{PeskinSchroeder}). They moreover satisfy the Jacobi identity
\begin{equation}\label{Jacobi}
   f^{ads} f^{bcs} + f^{bds} f^{cas} + f^{cds} f^{abs} = 0 .
\end{equation}
This identity is repeatedly needed in analyzing the gauge transformation behavior of the classical and quantum Yang-Mills equations.

Let us consider gauge transformations, which are given by (see, e.g., pp.\,490f of \cite{PeskinSchroeder} or Section 15.1 of \cite{WeinbergQFT2})
\begin{equation}\label{gaugetransA}
   A^a_\mu \rightarrow A^a_\mu + \partial_\mu \Lambda^a - g f^{abc} A^b_\mu \Lambda^c .
\end{equation}
Unlike for the Abelian case, the resulting transformation
\begin{equation}\label{gaugetransF}
   F^a_{\mu\nu} \rightarrow F^a_{\mu\nu} - g f^{abc} F^b_{\mu\nu} \Lambda^c + O(\Lambda^2) ,
\end{equation}
implies that the fields are not gauge invariant for non-Abelian Yang-Mills theories. However, by considering the combined transformation law
\begin{eqnarray}
   &\partial^\mu F^a_{\mu\nu} - g f^{abc} A^{b \mu} F^c_{\mu\nu} \rightarrow
   \partial^\mu F^a_{\mu\nu} - g f^{abc} A^{b \mu} F^c_{\mu\nu}& \nonumber \\
   &- g f^{abc} \big( \partial^\mu F^b_{\mu\nu}
   - g f^{bde} A^{d \mu} F^e_{\mu\nu} \big)  \Lambda^c + O(\Lambda^2) ,&
\label{gaugetransevol}
\end{eqnarray}
we realize that the field equations (\ref{classYMevol}) in the absence of external currents are gauge invariant. Moreover, the gauge transformation behavior of external currents required to obtain gauge invariant field equations becomes evident,
\begin{equation}\label{gaugetransJ}
   J^a_\mu \rightarrow J^a_\mu - g f^{abc} J^b_\mu \Lambda^c .
\end{equation}
For the gauge bosons, we here impose the covariant Lorenz gauge condition
\begin{equation}\label{Lorenzgauge}
   \partial_\mu A^{a \mu} = 0.
\end{equation}

We finally rewrite the Yang-Mills equations in a particular inertial system in the Maxwellian form which, in contrast to all the above equations, is no longer manifestly Lorentz covariant. Such a reformulation is a straightforward exercise (see, e.g., \cite{JackiwRossi80,SanchezQuimbay10} with some deviations in the choice of signs).

The counterparts of electric and magnetic fields are obtained by the convention
\begin{equation}\label{EBconvention}
   (F^a_{\mu\nu}) = \left(
                      \begin{array}{cccc}
                        0 & -E^a_1 & -E^a_2 & -E^a_3 \\
                        E^a_1 & 0 & B^a_3 & -B^a_2 \\
                        E^a_2 & -B^a_3 & 0 & B^a_1 \\
                        E^a_3 & B^a_2 & -B^a_1 & 0 \\
                      \end{array}
                    \right) .
\end{equation}
As pointed out before, for the non-Abelian case, these fields are not gauge invariant,
\begin{equation}\label{gaugetransB}
   \bm{B}^a \rightarrow \bm{B}^a - g f^{abc} \bm{B}^b \Lambda^c + O(\Lambda^2) ,
\end{equation}
and
\begin{equation}\label{gaugetransE}
   \bm{E}^a \rightarrow \bm{E}^a - g f^{abc} \bm{E}^b \Lambda^c + O(\Lambda^2) .
\end{equation}
For convenience, we define the additional component
\begin{equation}\label{E0convention}
   E^a_0 = \partial_\mu A^{a \mu} ,
\end{equation}
so that the Lorenz gauge condition (\ref{Lorenzgauge}) can be expressed as $E^a_0 = 0$.

The gauge-dependent field definitions (\ref{classYMfields}) become
\begin{equation}\label{classYMfields2}
   \bm{B}^a = \bm{\nabla} \times \bm{A}^a
   - \frac{1}{2} g f^{abc} \bm{A}^b \times \bm{A}^c ,
\end{equation}
\begin{equation}\label{classYMfields1}
   \bm{E}^a = \bm{\nabla} A^a_0 - \frac{\partial \bm{A}^a}{\partial t}
   - g f^{abc} \bm{A}^b A^c_0 ,
\end{equation}
and the field equations (\ref{classYMevol}) can be written as
\begin{equation}\label{classYMevol1}
   \bm{\nabla} \cdot \bm{E}^a -  g f^{abc} \bm{A}^b \cdot \bm{E}^c = -J^a_0 = J^{a 0} ,
\end{equation}
\begin{equation}\label{classYMevol2}
   \frac{\partial \bm{E}^a}{\partial t} - \bm{\nabla} \times \bm{B}^a
   - g f^{abc} ( A^b_0 \bm{E}^c - \bm{A}^b \times \bm{B}^c ) = - \bm{J}^a .
\end{equation}
Equations (\ref{classYMfields2})--(\ref{classYMevol2}) correspond to Eqs.~(3.15), (4.2c), (4.2a) and (4.2b) of \cite{JackiwRossi80}. In Eqs.~(\ref{classYMevol1}) and (\ref{classYMevol2}), we recognize the temporal and spatial components of the current $-g f^{abc} A^{b \mu} F^c_{\mu\nu}$ associated with the fields,
\begin{equation}\label{classfieldcurr0}
    \tilde{J}^{a 0} = g f^{abc} \bm{A}^b \cdot \bm{E}^c ,
\end{equation}
\begin{equation}\label{classfieldcurrj}
    \tilde{\bm{J}}^a = - g f^{abc} ( A^b_0 \bm{E}^c - \bm{A}^b \times \bm{B}^c ) .
\end{equation}

Equation (\ref{classYMfields2}) can be used to eliminate $\bm{B}^a$ and $\bm{B}^c$ from Eq.~(\ref{classYMevol2}). After eliminating $\bm{B}$ from the picture, we have the evolution equations
\begin{equation}\label{classYMevolA}
   \frac{\partial \bm{A}^a}{\partial t} = - \bm{E}^a + \bm{\nabla} A^a_0
   - g f^{abc} \bm{A}^b A^c_0 ,
\end{equation}
and
\begin{eqnarray}
   \frac{\partial \bm{E}^a}{\partial t} &=& - \bm{J}^a - \nabla^2 \bm{A}^a
   + \bm{\nabla} \bm{\nabla} \cdot \bm{A}^a
   + g f^{abc} A^b_0 \, \bm{E}^c \nonumber\\
   &+& g f^{abc} \, [ 2 \, \bm{A}^b \cdot \bm{\nabla} \bm{A}^c
   - \bm{A}^b \, \bm{\nabla} \cdot \bm{A}^c
   + (\bm{\nabla} \bm{A}^b) \cdot \bm{A}^c ] \nonumber\\
   &-& g^2 f^{abs} f^{cds} \bm{A}^b \cdot \bm{A}^c \, \bm{A}^d ,
\label{classYMevolE}
\end{eqnarray}
together with the relation (\ref{classYMevol1}). These equations define classical Yang-Mills theories in a formulation that does not manifestly exhibit Lorentz or gauge symmetry. These equations imply a conservation law with source term,
\begin{equation}\label{classconti}
   \frac{\partial \tilde{J}^{a 0}}{\partial t} + \bm{\nabla} \cdot \tilde{\bm{J}}^a =
   - g f^{abc} \, A^b_\mu J^{c \mu} .
\end{equation}

A wave equation for $\bm{A}^a$ is obtained by subtracting Eq.~(\ref{classYMevol2}) from the time derivative of Eq.~(\ref{classYMfields1}). Similarly, a the wave equation for $A^{a 0}$ is obtained by subtracting Eq.~(\ref{classYMevol1}) from the divergence of Eq.~(\ref{classYMfields1}). We thus find the wave equations
\begin{eqnarray}
   \left( \frac{\partial^2}{\partial t^2} - \nabla^2 \right) A^a_0 &=&
   J^a_0 + g f^{abc} \, A^{b \mu} ( \partial_0 A^c_\mu - 2 \partial_\mu A^c_0)
   \nonumber \\
   &+& g^2 f^{abs} f^{cds} \bm{A}^b \cdot \bm{A}^c \, A^d_0
   \nonumber \\
   &-& ( \delta^{ac} \partial_0 - g f^{abc} \, A^b_0 ) \, \partial_\mu A^{c \mu} ,
\label{classYMevol1w}
\end{eqnarray}
and
\begin{eqnarray}
   \left( \frac{\partial^2}{\partial t^2} - \nabla^2 \right) \bm{A}^a &=&
   \bm{J}^a + g f^{abc} \, A^{b \mu} ( \bm{\nabla} A^c_\mu - 2 \partial_\mu \bm{A}^c)
   \nonumber \\
   &+& g^2 f^{abs} f^{cds} A^{b \mu} A^c_\mu \, \bm{A}^d
   \nonumber \\
   &-& ( \delta^{ac} \bm{\nabla} - g f^{abc} \, \bm{A}^b ) \, \partial_\mu A^{c \mu} ,
\label{classYMevol2w}
\end{eqnarray}
in each of which the last term disappears for the Lorenz gauge (\ref{Lorenzgauge}) (cf.\ appendix of \cite{SanchezQuimbay10}). A closer look at Eqs.~(\ref{classYMevol1w}) and (\ref{classYMevol2w}) reveals that one recovers Lorentz covariance in these wave equations.

\section{Fock space}\label{secFockspace}
We introduce a Fock space together with a collection of adjoint operators $a^{a \alpha\,\dag}_{\bm{q}}$ and $a^{a \alpha}_{\bm{q}}$ creating and annihilating field quanta, such as photons or gluons, with momentum $\bm{q} \in \bar{K}^3_\times$ and polarization state $\alpha \in \{0,1,2,3\}$. The space of momentum vectors is given by the simple cubic lattice
\begin{equation}\label{K3latticedef}
    \bar{K}^3_\times = \left\{ \bm{q} =  K_L \, (z_1,z_2,z_3) \, |
    \, z_1,z_2,z_3 \in \mathbb{Z} \right\} ,
\end{equation}
where $K_L$ is a lattice constant in momentum space, which is assumed to be small because it is given by the inverse system size. The additional label $a$ is associated with the infinitesimal generators of an underlying Lie group [assuming $3$ values for $SU(2)$ corresponding to the W$^+$, W$^-$, and Z$^0$ bosons mediating weak interactions, $8$ values for $SU(3)$ corresponding to the gluon ``color octet'' mediating strong interactions, and $N^2-1$ values for general $SU(N)$]. For simplicity, we occasionally refer to the gauge or vector bosons associated with the Yang-Mills field as gluons, for which $a$ labels eight different ``color'' states. The gluon creation and annihilation operators satisfy canonical commutation relations.

Details on the construction of Fock spaces can be found, e.g., in Secs.~1 and 2 of \cite{FetterWalecka}, in Secs.~12.1 and 12.2 of \cite{BjorkenDrell}, or in Sec.~1.2.1 of \cite{hcoqft}. We assume that the collection of the states created by multiple application of all the above creation operators on a ground state, which is annihilated by all the corresponding annihilation operators, is complete.

In the following, we repeatedly need the properties of the polarization states. We hence give our explicit representations. The temporal unit four-vector,
\begin{equation}\label{polarization0}
    (\bar{n}^0_{\mu \bm{q}}) =
    \left( \begin{array}{c}
      1 \\
      0 \\
      0 \\
      0 \\
    \end{array} \right) ,
\end{equation}
is actually independent of $\bm{q}$. The three orthonormal spatial \index{Polarization}polarization vectors are chosen as
\begin{equation}\label{polarization1}
    (\bar{n}^1_{\mu \bm{q}}) = \frac{1}{\sqrt{q_1^2+q_2^2}}
    \left( \begin{array}{c}
      0 \\
      q_2 \\
      - q_1 \\
      0 \\
    \end{array} \right) ,
\end{equation}
\begin{equation}\label{polarization2}
    (\bar{n}^2_{\mu \bm{q}}) = \frac{1}{q \sqrt{q_1^2+q_2^2}}
    \left( \begin{array}{c}
      0 \\
      q_1 q_3 \\
      q_2 q_3\\
      - q_1^2 - q_2^2 \\
    \end{array} \right) ,
\end{equation}
and
\begin{equation}\label{polarization3}
    (\bar{n}^3_{\mu \bm{q}}) = \frac{1}{q}
    \left( \begin{array}{c}
      0 \\
      q_1\\
      q_2\\
      q_3\\
    \end{array} \right) ,
\end{equation}
where $q=|\bm{q}|$. The polarization vectors $\bar{n}^1_{\bm{q}}$ and $\bar{n}^2_{\bm{q}}$ correspond to transverse gluons, $\bar{n}^3_{\bm{q}}$ corresponds to longitudinal gluons. Note the symmetry property
\begin{equation}\label{polarsym}
    \bar{n}^\alpha_{-\bm{q}} = (-1)^\alpha \, \bar{n}^\alpha_{\bm{q}} \,.
\end{equation}

In the BRST approach, one introduces the additional pairs $B^{a\, \dag}_{\bm{q}}$, $B^{a}_{\bm{q}}$ and $D^{a\, \dag}_{\bm{q}}$, $D^{a}_{\bm{q}}$ of creation and annihilation operators associated with ghost particles and their antiparticles. They are assumed to satisfy canonical anticommutation relations.

Fock spaces come with canonical inner products, $s^{\rm can}$. The superscript $\dag$ on the creation operators can actually be interpreted as the adjoint with respect to the canonical inner product. In gauge theories, the canonical inner product is not the physical one. We need an additional signed inner product, $s^{\rm sign}$, for which states with negative norm exist (so that it is not an inner product in a strict sense). In both inner products, the natural base vectors of a Fock space, which are characterized by the occupation numbers of the various particles in the system, are orthogonal. If the canonical norm of a Fock base vector is one, the signed norm of that vector is obtained by introducing a factor $-1$ for every temporal gauge boson and for every ghost particle created by $B^{a\, \dag}_{\bm{q}}$. The adjoint of an operator with respect to the signed inner product is indicated by the superscript $\ddag$. The commutation and signed product properties of all particles are compiled in Table~\ref{tablecreationprop}.

\begin{table}
\caption[ ]{Properties associated with the various creation operators used for constructing the Fock space}
\renewcommand{\arraystretch}{1.5}
\begin{center}
\begin{tabular}{|c|c|c|c|}
\hline
\multicolumn{2}{|c|}{commuting} & \multicolumn{2}{c|}{anticommuting} \\
\hline
\ - metric \ & \ + metric \ & \ - metric \ & \ + metric \ \\
\hline
$a^{a 0\, \dag}_{\bm{q}}$ & $a^{a j\, \dag}_{\bm{q}}$ & $B^{a\, \dag}_{\bm{q}}$ & $D^{a\, \dag}_{\bm{q}}$ \\
\hline
\end{tabular}
\end{center}
\renewcommand{\arraystretch}{1}
\label{tablecreationprop}
\end{table}

\section{Basics of BRST approach}
A brief discussion of BRST quantization in terms of creation and annihilation operators can be found on pp.\,239--240 of \cite{Nemeschanskyetal86}. For quantum electrodynamics, all the details have been elaborated in Section~3.2.5 of \cite{hcoqft}. We here generalize these ideas to non-Abelian gauge theories.

\subsection{Ghost particle operators}
In the usual interpretation, the operators $B^{a\, \dag}_{\bm{q}}$ and $D^{a\, \dag}_{\bm{q}}$ in Table~\ref{tablecreationprop} create massless ghost particles and their antiparticles. The Fourier modes of the corresponding fields are hence given by
\begin{equation}\label{ghostfields}
    c^a_{\bm{q}} = \frac{1}{\sqrt{2q}} (D^{a\, \dag}_{\bm{q}} - B^a_{-\bm{q}} ) ,
    \quad
    \bar{c}^a_{\bm{q}} = \frac{1}{\sqrt{2q}} (B^{a\, \dag}_{\bm{q}} + D^a_{-\bm{q}} ) .
\end{equation}
If we further define the fields
\begin{equation}\label{ghostfieldsdot}
    \dot{c}^a_{\bm{q}} = \iR \sqrt{\frac{q}{2}} (D^{a\, \dag}_{\bm{q}} + B^a_{-\bm{q}} ) ,
    \quad
    \dot{\bar{c}}^a_{\bm{q}} = \iR \sqrt{\frac{q}{2}} (B^{a\, \dag}_{\bm{q}} - D^a_{-\bm{q}} ) ,
\end{equation}
the only non-vanishing anticommutation relations among the operators introduced in Eqs.~(\ref{ghostfields}) and (\ref{ghostfieldsdot}) are given by
\begin{equation}\label{ccbardotanticomrel}
   \Qantico{c^a_{\bm{q}}}{\dot{\bar{c}}^b_{\bm{q}'}} =
   - \iR \delta_{ab} \delta_{\bm{q}+\bm{q}',\bm{0}} , \quad
   \Qantico{\bar{c}^a_{\bm{q}}}{\dot{c}^b_{\bm{q}'}} =
   \iR \delta_{ab} \delta_{\bm{q}+\bm{q}',\bm{0}} .
\end{equation}
Further note the adjointness properties with respect to the signed inner product,
\begin{equation}\label{cadjoint}
   c^{a \, \ddag}_{\bm{q}} = \bar{c}^a_{-\bm{q}} , \qquad
   \dot{c}^{a \, \ddag}_{\bm{q}} = \dot{\bar{c}}^a_{-\bm{q}} ,
\end{equation}
and the simple transformation rule for the energy of noninteracting massless ghosts (neglecting an irrelevant constant to achieve normal ordering),
\begin{equation}\label{ghostenergytransf}
    \sum_{\bm{q} \in \bar{K}^3_\times} q
    \big(  B^{a \, \dag}_{\bm{q}} B^a_{\bm{q}} + D^{a \, \dag}_{\bm{q}} D^a_{\bm{q}} \big) =
    \sum_{\bm{q} \in \bar{K}^3_\times}
    \big( \dot{c}^a_{\bm{q}} \dot{\bar{c}}^a_{-\bm{q}} + q^2 c^a_{\bm{q}} \bar{c}^a_{-\bm{q}} \big) .
\end{equation}

At this point a comment on notation should be useful. One might be tempted to interpret the dots in the symbols $\dot{c}^a_{\bm{q}}$, $\dot{\bar{c}}^a_{\bm{q}}$ introduced in Eq.~(\ref{ghostfieldsdot}) as time derivatives. However, this interpretation works only for massless free ghost particles. For quantum electrodynamics, this interpretation would actually be justified. For non-Abelian gauge theories, however, the ghost particles are no longer free and $\dot{c}^a_{\bm{q}}$, $\dot{\bar{c}}^a_{\bm{q}}$ should be recognized as nothing but the operators defined in Eq.~(\ref{ghostfieldsdot}); they do not coincide with time derivatives that could be defined in terms of the full Hamiltonian.

\subsection{Alternative representation of ghosts}
It has been pointed out by Kugo and Ojima \cite{KugoOjima78a} that the lack of self-adjointness of the ghost operators expressed in Eq.~(\ref{cadjoint}) is a serious disadvantage in constructing the physical states. In non-Abelian Yang-Mills theories, it moreover keeps the Hamiltonian from being self-adjoint. Those authors hence recommend an alternative representation of ghosts that, however, requires non-canonical anticommutation relations for the creation and annihilation operators $B^{a\, \dag}_{\bm{q}}$, $B^{a}_{\bm{q}}$ and $D^{a\, \dag}_{\bm{q}}$, $D^{a}_{\bm{q}}$. In order to base our development on a well-defined Fock space, we strongly prefer to keep canonical anticommutation relations. We therefore propose a new representation of the ghost fields in terms of two creation and two annihilation operators,
\begin{equation}\label{ghostfield1}
    c^a_{\bm{q}} = \frac{1}{2\sqrt{q}}
    ( B^{a\, \dag}_{\bm{q}} + D^{a\, \dag}_{\bm{q}} - B^a_{-\bm{q}} + D^a_{-\bm{q}} ) ,
\end{equation}
\begin{equation}\label{ghostfield2}
    \bar{c}^a_{\bm{q}} = \frac{1}{2\sqrt{q}}
    ( B^{a\, \dag}_{\bm{q}} - D^{a\, \dag}_{\bm{q}} + B^a_{-\bm{q}} + D^a_{-\bm{q}} ) ,
\end{equation}
\begin{equation}\label{ghostfielddot1}
    \dot{c}^a_{\bm{q}} = \frac{\iR\sqrt{q}}{2}
    ( B^{a\, \dag}_{\bm{q}} + D^{a\, \dag}_{\bm{q}} + B^a_{-\bm{q}} - D^a_{-\bm{q}} ) ,
\end{equation}
and
\begin{equation}\label{ghostfielddot2}
    \dot{\bar{c}}^a_{\bm{q}} = \frac{\iR\sqrt{q}}{2}
    ( B^{a\, \dag}_{\bm{q}} - D^{a\, \dag}_{\bm{q}} - B^a_{-\bm{q}} - D^a_{-\bm{q}} ) .
\end{equation}
A straightforward calculation shows that the resulting anticommutation relations for the operators defined in Eqs.~(\ref{ghostfield1})--(\ref{ghostfielddot2}) are identical to the previously found ones, where the only nontrivial anticommutators are given in Eq.~(\ref{ccbardotanticomrel}). Therefore, also most of the subsequent calculations for the two ways of introducing ghost fields are identical. Moreover, Eq.~(\ref{ghostenergytransf}) remains valid.

The big advantage of the new definitions are the self-adjointness properties
\begin{equation}\label{cadjointc}
   c^{a \, \ddag}_{\bm{q}} = c^a_{-\bm{q}} , \quad
   \dot{c}^{a \, \ddag}_{\bm{q}} = \dot{c}^a_{-\bm{q}} ,\quad
   \bar{c}^{a \, \ddag}_{\bm{q}} = - \bar{c}^a_{-\bm{q}} , \quad
   \dot{\bar{c}}^{a \, \ddag}_{\bm{q}} = - \dot{\bar{c}}^a_{-\bm{q}} .
\end{equation}
These properties imply the self-adjointness of products such as
\begin{equation}\label{cadjointcc}
   \big( c^a_{\bm{q}} \bar{c}^b_{\bm{q}'} \big)^\ddag = c^a_{-\bm{q}} \bar{c}^b_{-\bm{q}'} .
\end{equation}
The systematic occurrence of $c\bar{c}$ pairs is attractive also from another viewpoint: We obtain a symmetry under rescaling $c$ and $\bar{c}$ by inverse factors and hence an additional conserved quantity in the ghost domain (see Section~\ref{secaddconscharge} for details).
In view of Eqs.~(\ref{cadjoint}) and (\ref{cadjointc}), we refer to the two options for introducing ghost operators as ``cross-adjointness'' (option 1) and ``self-adjointness'' (option 2), respectively. We once more emphasize that both options are realized on exactly the same Fock space.

\subsection{BRST charge and transformations}
To keep track of sums and differences of photon creation and annihilation operators, to evaluate commutators and anticommutators in an efficient manner, and to facilitate the comparison with the classical theory, we introduce the operators
\begin{eqnarray}
    \alpha^a_{\bm{q}} &=& \frac{1}{\sqrt{2q}} \, \Big[
    \bar{n}^0_{\bm{q}} ( a^{a 0 \, \dag}_{\bm{q}} - a^{a 0}_{-\bm{q}} )
    + \bar{n}^1_{\bm{q}} ( a^{a 1 \, \dag}_{\bm{q}} - a^{a 1}_{-\bm{q}} )
    \nonumber \\
    &+& \bar{n}^2_{\bm{q}} ( a^{a 2 \, \dag}_{\bm{q}} + a^{a 2}_{-\bm{q}} )
    + \bar{n}^3_{\bm{q}} ( a^{a 3 \, \dag}_{\bm{q}} - a^{a 3}_{-\bm{q}} ) \Big] ,
\label{auxalphadef}
\end{eqnarray}
and
\begin{eqnarray}
    \varepsilon^a_{\bm{q}} &=& - \iR \sqrt{\frac{q}{2}} \, \Big[
    \bar{n}^0_{\bm{q}} ( a^{a 0 \, \dag}_{\bm{q}} + a^{a 0}_{-\bm{q}}
    + a^{a 3 \, \dag}_{\bm{q}} - a^{a 3}_{-\bm{q}} )
    \nonumber \\
    &+& \bar{n}^1_{\bm{q}} ( a^{a 1 \, \dag}_{\bm{q}} + a^{a 1}_{-\bm{q}} )
    + \bar{n}^2_{\bm{q}} ( a^{a 2 \, \dag}_{\bm{q}} - a^{a 2}_{-\bm{q}} ) \nonumber \\
    &+& \bar{n}^3_{\bm{q}} ( a^{a 3 \, \dag}_{\bm{q}} + a^{a 3}_{-\bm{q}}
    + a^{a 0 \, \dag}_{\bm{q}} - a^{a 0}_{-\bm{q}} ) \Big] .
\label{auxetadef}
\end{eqnarray}
% The inverse relations are given in Appendix~\ref{appphotoninv}.
According to Eqs.~(3.73) and (3.81) of \cite{hcoqft}, $\alpha^a_{\bm{q}}$ corresponds to the four-vector potential and $\varepsilon^a_{\bm{q}}$ to the electric field [augmented by the time component introduced in Eq.~(\ref{E0convention})]. These four-vectors satisfy canonical commutation relations
\begin{equation}\label{cancomrelaleta}
   \Qcommu{\varepsilon^{a \mu}_{\bm{p}}}{\alpha^{b \nu}_{\bm{q}}} =
   \iR \eta^{\mu\nu} \delta_{ab} \delta_{\bm{p}+\bm{q},\bm{0}} ,
\end{equation}
\begin{equation}\label{cancomrelaleta0}
   \Qcommu{\alpha^{a \mu}_{\bm{p}}}{\alpha^{b \nu}_{\bm{q}}} =
   \Qcommu{\varepsilon^{a \mu}_{\bm{p}}}{\varepsilon^{b \nu}_{\bm{q}}} = 0 ,
\end{equation}
where $\eta^{\mu\nu}$ represents the Minkowski metric, and possess the following self-adjointness properties with respect to the signed inner product,
\begin{equation}\label{aletaadjoint}
   \alpha^{a \, \ddag}_{\bm{q}} = \alpha^a_{-\bm{q}} , \qquad
   \varepsilon^{a \, \ddag}_{\bm{q}} = \varepsilon^a_{-\bm{q}} .
\end{equation}

The energy of temporal and longitudinal noninteracting vector bosons can be written as
\begin{eqnarray}
    \sum_{\bm{q} \in \bar{K}^3_\times} q
    \big( a^{a 0 \, \dag}_{\bm{q}} a^{a 0}_{\bm{q}}
    + a^{a 3 \, \dag}_{\bm{q}} a^{a 3}_{\bm{q}} \big) &=&
    \nonumber\\
    && \hspace{-10em} \sum_{\bm{q} \in \bar{K}^3_\times}
    \iR q_j \big( \varepsilon^{a j}_{\bm{q}} \, \alpha^{a 0}_{-\bm{q}}
    + \alpha^{a j}_{\bm{q}} \, \varepsilon^{a 0}_{-\bm{q}} \big)
    \nonumber\\
    && \hspace{-10em} + \, \frac{1}{2} \sum_{\bm{q} \in \bar{K}^3_\times}
    \left( \frac{q_j q_k}{q^2} \, \varepsilon^{a j}_{\bm{q}} \, \varepsilon^{a k}_{-\bm{q}}
    - \varepsilon^{a 0}_{\bm{q}} \, \varepsilon^{a 0}_{-\bm{q}} \right) . \qquad
\label{templongenergytransf}
\end{eqnarray}

In the following, we use the notation
\begin{equation}\label{convolution}
  (A B)_{\bm{q}} = \frac{1}{\sqrt{V}} \sum_{\bm{q}',\bm{q}'' \in \bar{K}^3_\times}
  \delta_{\bm{q}'+\bm{q}'',\bm{q}} \, A_{\bm{q}'} B_{\bm{q}''} ,
\end{equation}
for any two $\bm{q}$ dependent operators $A$ and $B$. This notation for convolutions allows us to obtain more compact equations, where working on the infinite lattice $\bar{K}^3_\times$ (that is, in the thermodynamic limit) is crucial for obtaining the usual properties of convolutions (the relevance of high momenta for a symmetry involving derivatives has been pointed out on p.\,169 of \cite{hcoqft}; on a finite lattice, rigorous BRST symmetry cannot be expected).

The following expression for the BRST charge is inspired, for example, by Eq.~(5.10) of \cite{Nemeschanskyetal86} or Eq.~(2.22) of \cite{KugoOjima78b}, but the proper formulation in the present setting is not entirely straightforward,
\begin{eqnarray}
    Q &=& \sum_{\bm{q} \in \bar{K}^3_\times}
    \big( \varepsilon^a_{0 \bm{q}} \, \dot{c}^a_{-\bm{q}} -
    \iR q_j \, \varepsilon^{a j}_{\bm{q}} \, c^a_{-\bm{q}} \big) \nonumber\\
    &+& g f^{abc} \sum_{\bm{q} \in \bar{K}^3_\times}
    \Big( \varepsilon^{a \mu} \alpha^b_\mu - \dot{c}^a \bar{c}^b
    + \frac{1}{2} c^a \dot{\bar{c}}^b \Big)_{\bm{q}} \, c^c_{-\bm{q}} . \qquad
\label{BRSTcharge}
\end{eqnarray}
Rearranging the factors in the three-particle collision terms proportional to $g$ in Eq.~(\ref{BRSTcharge}) is unproblematic because, whenever a nonzero contribution might arise from a nontrivial anticommutation relation, the corresponding structure constant with two equal labels vanishes. In particular, normal ordering is not an issue in three-particle collision terms, which is an appealing feature of such interactions.

The BRST charge (\ref{BRSTcharge}) implies the following characteristic BRST transformations for non-Abelian gauge theories,
\begin{eqnarray}
  \delta \alpha^a_{0 \bm{q}} &=& \dot{c}^a_{\bm{q}} -
  g f^{abc} \, (\alpha^b_0 c^c)_{\bm{q}} ,
  \nonumber\\
  \delta \alpha^a_{j \bm{q}} &=& -\iR q_j \, c^a_{\bm{q}} -
  g f^{abc} (\alpha^b_j c^c)_{\bm{q}} ,
  \nonumber\\
  \delta \varepsilon^a_{\mu \bm{q}} &=& - g f^{abc} \, (\varepsilon^b_\mu c^c)_{\bm{q}} .
  \nonumber\\
  \delta c^a_{\bm{q}} &=& \frac{1}{2} \, g f^{abc} \, (c^b c^c)_{\bm{q}} ,
  \nonumber\\
  \delta \dot{c}^a_{\bm{q}} &=& g f^{abc} \, (\dot{c}^b c^c)_{\bm{q}} ,
  \nonumber\\
  \delta \bar{c}^a_{\bm{q}} &=& \varepsilon^{a 0}_{\bm{q}} +
  g f^{abc} \, ( \bar{c}^b c^c )_{\bm{q}} ,
  \nonumber\\
  \delta \dot{\bar{c}}^a_{\bm{q}} &=&
  - \iR q_j \, \varepsilon^{a j}_{\bm{q}} +
  g f^{abc} \, ( \varepsilon^{b \mu} \alpha^c_\mu
  - \dot{c}^b \bar{c}^c + c^b \dot{\bar{c}}^c )_{\bm{q}} , \qquad
\label{BRSTtransform}
\end{eqnarray}
where $\delta\cdot = \iR \Qcommu{Q}{\cdot}$ for boson and $\delta\cdot = \iR \Qantico{Q}{\cdot}$ for fermion operators. The first two lines of Eq.~(\ref{BRSTtransform}) correspond to the gauge transformation (\ref{gaugetransA}), the third line of Eq.~(\ref{BRSTtransform}) corresponds to the transformation (\ref{gaugetransE}).

With the BRST charge (\ref{BRSTcharge}) we have the tool to discuss all aspects of BRST symmetry. In particular, we can construct a BRST invariant Hamiltonian and the physical states in the large Fock space involving ghosts. The compatibility of BRST invariance with Lorentz symmetry is visible in the first two lines of Eq.~(\ref{BRSTtransform}), which relates the BRST transformations of the four-vector potential to the time and space derivatives of the ghost field $c$. The nilpotency of the BRST charge ($Q^2=0$), which is crucial for the handling of BRST symmetry, is verified in Appendix~\ref{appnilpotent}. As a somewhat simpler alternative, one can verify $\delta Q = 0$ by means of the BRST transformations (\ref{BRSTtransform}). The first two lines of Eq.~(\ref{BRSTtransform}), which express the essence of the classical gauge transformations (\ref{gaugetransA}), and the nilpotency of $Q$ provide the deeper reasons for writing the BRST charge in the form given in Eq.~(\ref{BRSTcharge}).

\section{Construction of BRST invariant operators}\label{secopconstruction}
A simple way of constructing BRST invariant operators is based on the identity
\begin{equation}\label{invariantopident}
   \Qcommu{Q}{\iR\Qantico{Q}{X}} = 0 ,
\end{equation}
which, for any operator $X$, follows trivially from the nilpotency of $Q$. In other words, any operator $\iR\Qantico{Q}{X}$ is BRST invariant as it commutes with $Q$. In practice, one chooses $X$ to produce a desirable term and one automatically gets all the additional terms required for BRST invariance. We illustrate the idea for some simple choices of $X$, which consist of an $\alpha$ or $\varepsilon$ paired with a $c$ or $\bar{c}$.

To produce terms of the type $\dot{c} \dot{\bar{c}}$, we choose
\begin{equation}\label{BRSTinvop1x}
   X_1 = \sum_{\bm{q} \in \bar{K}^3_\times} \alpha^a_{0 \bm{q}} \, \dot{\bar{c}}^a_{-\bm{q}} .
\end{equation}
Straightforward calculations based on the product rule and the results in Eq.~(\ref{BRSTtransform}) give
\begin{eqnarray}
   \iR\Qantico{Q}{X_1} &=& \sum_{\bm{q} \in \bar{K}^3_\times}
   \big( \dot{c}^a_{\bm{q}} \dot{\bar{c}}^a_{-\bm{q}} +
   \iR q_j \, \varepsilon^{a j}_{\bm{q}} \, \alpha^{a 0}_{-\bm{q}} \big)
   \nonumber\\
   &+& g f^{abc} \sum_{\bm{q} \in \bar{K}^3_\times}
   \big( \varepsilon^{a j} \alpha^b_j - \dot{c}^a \bar{c}^b \big)_{-\bm{q}} \alpha^c_{0 \bm{q}} .
   \qquad
\label{BRSTinvop1}
\end{eqnarray}
Similarly, to produce terms of the type $c \bar{c}$, we choose
\begin{equation}\label{BRSTinvop2x}
   X_2 = \sum_{\bm{q} \in \bar{K}^3_\times} \iR q_j \, \alpha^a_{j \bm{q}} \, \bar{c}^a_{-\bm{q}} ,
\end{equation}
and we obtain the BRST invariant operator
\begin{eqnarray}
   \iR\Qantico{Q}{X_2} &=& \sum_{\bm{q} \in \bar{K}^3_\times}
   \big( q^2 c^a_{\bm{q}} \bar{c}^a_{-\bm{q}} +
   \iR q_j \, \alpha^{a j}_{\bm{q}} \, \varepsilon^{a 0}_{-\bm{q}} \big)
   \nonumber\\
   &-& g f^{abc} \sum_{\bm{q} \in \bar{K}^3_\times}
   \iR q_j \, c^a_{\bm{q}} (\bar{c}^b \alpha^c)_{-\bm{q}} .
\label{BRSTinvop2}
\end{eqnarray}
Another interesting choice is given by
\begin{equation}\label{BRSTinvop3x}
   X_3 = \sum_{\bm{q} \in \bar{K}^3_\times} \varepsilon^a_{0 \bm{q}} \, \bar{c}^a_{-\bm{q}} .
\end{equation}
It leads to the simple BRST invariant operator
\begin{equation}\label{BRSTinvop3}
   \iR\Qantico{Q}{X_3} = - \sum_{\bm{q} \in \bar{K}^3_\times}
   \varepsilon^{a 0}_{\bm{q}} \, \varepsilon^{a 0}_{-\bm{q}} .
\end{equation}

In summary, we have used bilinear operators $X$ to produce a number of BRST invariant operators $\iR\Qantico{Q}{X}$. A comparison with Eqs.~(\ref{ghostenergytransf}) and (\ref{templongenergytransf}) shows that, for $g=0$, all our examples contain contributions from the free Hamiltonian. This observation is very useful for the subsequent construction of a BRST invariant Hamiltonian.

\section{Yang-Mills Hamiltonian}
We now construct a BRST invariant Hamiltonian and discuss some implications. The construction is based entirely on symmetry considerations.

\subsection{Construction of Hamiltonian}
We start from the energy of the noninteracting transverse polarizations of the vector bosons,
\begin{eqnarray}
    &&\sum_{\bm{q} \in \bar{K}^3_\times} q
    \big( a^{a 1 \, \dag}_{\bm{q}} a^{a 1}_{\bm{q}}
    + a^{a 2 \, \dag}_{\bm{q}} a^{a 2}_{\bm{q}} \big) \nonumber\\
    && = \, \frac{1}{2} \sum_{\bm{q} \in \bar{K}^3_\times}
    \big( \bar{n}^1_{j \bm{q}} \bar{n}^1_{k \bm{q}}
    + \bar{n}^2_{j \bm{q}} \bar{n}^2_{k \bm{q}} \big)
    \big( q^2 \alpha^{a j}_{\bm{q}} \alpha^{a k}_{-\bm{q}}
    + \varepsilon^{a j}_{\bm{q}} \varepsilon^{a k}_{-\bm{q}} \big) . \nonumber\\
    &&
\label{Hfreetrv}
\end{eqnarray}
A more convenient starting point is actually given by
\begin{eqnarray}
    \Phi &=& \sum_{\bm{q} \in \bar{K}^3_\times} q
    \big( a^{a 1 \, \dag}_{\bm{q}} a^{a 1}_{\bm{q}}
    + a^{a 2 \, \dag}_{\bm{q}} a^{a 2}_{\bm{q}} \big)
    +\frac{1}{2} \sum_{\bm{q} \in \bar{K}^3_\times}
    \frac{q_j q_k}{q^2} \, \varepsilon^{a j}_{\bm{q}} \, \varepsilon^{a k}_{-\bm{q}}
    \nonumber\\
    &=& \frac{1}{2} \sum_{\bm{q} \in \bar{K}^3_\times}
    \Big[ \big( q^2 \delta_{jk} - q_j q_k \big)
    \alpha^{a j}_{\bm{q}} \alpha^{a k}_{-\bm{q}}
    + \varepsilon^{a j}_{\bm{q}} \varepsilon^{a j}_{-\bm{q}} \Big] . \nonumber\\
    &&
\label{Hfreetrvx}
\end{eqnarray}
A straightforward calculation yields
\begin{equation}\label{transtranf}
   \iR\Qcommu{Q}{\Phi} = - g f^{abc} \sum_{\bm{q} \in \bar{K}^3_\times}
    ( q^2 \delta_{jk} - q_j q_k ) \, \alpha^{a j}_{\bm{q}} (\alpha^{b k} c^c)_{-\bm{q}} ,
\end{equation}
so that $\Phi$ is not yet BRST invariant for $g \neq 0$. As a compensating term we consider
\begin{equation}\label{transcomp}
   \Phi' = \frac{g f^{abc}}{\sqrt{V}} \sum_{\bm{q},\bm{q}',\bm{q}'' \in \bar{K}^3_\times}
   \delta_{\bm{q}+\bm{q}'+\bm{q}'',\bm{0}} \,
   (- \iR q_j) \, \alpha^a_{k \bm{q}} \alpha^b_{k \bm{q}'} \alpha^c_{j \bm{q}''} ,
\end{equation}
for which we find
\begin{eqnarray}
   \iR\Qcommu{Q}{\Phi'} &=& g f^{abc} \sum_{\bm{q} \in \bar{K}^3_\times}
   ( q^2 \delta_{jk} - q_j q_k ) \, \alpha^{a j}_{\bm{q}} (\alpha^{b k} c^c)_{-\bm{q}}
   \nonumber\\
   &+& \frac{g^2 f^{abs} f^{cds}}{V} \hspace{-1em}
   \sum_{\bm{q},\bm{q}',\bm{p},\bm{p}' \in \bar{K}^3_\times}
   \delta_{\bm{q}+\bm{q}'+\bm{p}+\bm{p}',\bm{0}} \nonumber\\
   && \hspace{6em} \times \, \iR q_j \, c^a_{\bm{q}} \, \alpha^b_{k \bm{q}'}
   \alpha^c_{j \bm{p}} \alpha^d_{k \bm{p}'} .
\label{transtranfc}
\end{eqnarray}
The term proportional to $g$ indeed cancels the contribution from $\Phi$ in Eq.~(\ref{transtranf}), however, a new second-order term in $g$ arises. This second-order term has been written in a compact form by arranging the three contributions resulting from the product rule in a standard form where, for rearranging the last contribution, the Jacobi identity is required. In this compact form we easily find the final compensating term
\begin{equation}\label{transcompp}
   \Phi'' = \frac{g^2 f^{abs} f^{cds}}{4 V} \hspace{-1em}
   \sum_{\bm{q},\bm{q}',\bm{p},\bm{p}' \in \bar{K}^3_\times}
   \delta_{\bm{q}+\bm{q}'+\bm{p}+\bm{p}',\bm{0}} \,
   \alpha^a_{j \bm{q}} \alpha^b_{k \bm{q}'} \alpha^c_{j \bm{p}} \alpha^d_{k \bm{p}'} ,
\end{equation}
in which all for spatial gauge boson operators are on an equal footing. In order to show that no leftover higher-order terms occur one needs to use the Jacobi identity. The operator $\Phi+\Phi'+\Phi''$ hence is BRST invariant.

The operator $\Phi$ contains the energy of the transverse gauge bosons and, according to Eq.~(\ref{templongenergytransf}), also part of energy of temporal and longitudinal gauge bosons. The missing parts are found in the BRST invariant operators established in Section~\ref{secopconstruction}. Moreover, these BRST invariant operators contain the energy (\ref{ghostenergytransf}) of noninteracting ghost particles. By collecting terms, we get the BRST invariant total Hamiltonian
\begin{equation}\label{Htotal}
   H = H_{\rm gb}^{\rm free} + H_{\rm gb}^{\rm coll} ,
\end{equation}
consisting of the energy of free massless gauge bosons and ghost particles
\begin{equation}\label{Hfreegb}
    H_{\rm gb}^{\rm free} = \sum_{\bm{q} \in \bar{K}^3_\times} q
    \Big( a^{a \alpha \, \dag}_{\bm{q}} a^{a \alpha}_{\bm{q}}
    + B^{a \, \dag}_{\bm{q}} B^a_{\bm{q}} + D^{a \, \dag}_{\bm{q}} D^a_{\bm{q}} \Big) ,
\end{equation}
and the interaction (confer to Eq.~(3) of \cite{Hayata12}),
\begin{eqnarray}
   H_{\rm gb}^{\rm coll} &=& \frac{g f^{abc}}{\sqrt{V}} \sum_{\bm{q},\bm{q}',\bm{q}'' \in \bar{K}^3_\times}
   \delta_{\bm{q}+\bm{q}'+\bm{q}'',\bm{0}} \, \Big(
   \varepsilon^{a j}_{\bm{q}} \alpha^b_{j \bm{q}'} \alpha^c_{0 \bm{q}''} \nonumber\\
   &-& \dot{c}^a_{\bm{q}} \bar{c}^b_{\bm{q}'} \alpha^c_{0 \bm{q}''}
   - \iR q_j \, \alpha^a_{k \bm{q}} \alpha^b_{k \bm{q}'} \alpha^c_{j \bm{q}''}
   - \iR q_j \, c^a_{\bm{q}} \bar{c}^b_{\bm{q}'} \alpha^c_{j \bm{q}''} \Big) \nonumber\\
   && \hspace{-2.5em} + \, \frac{g^2 f^{abs} f^{cds}}{4 V} \hspace{-1em}
   \sum_{\bm{q},\bm{q}',\bm{p},\bm{p}' \in \bar{K}^3_\times}
   \delta_{\bm{q}+\bm{q}'+\bm{p}+\bm{p}',\bm{0}} \,
   \alpha^a_{j \bm{q}} \alpha^b_{k \bm{q}'} \alpha^c_{j \bm{p}} \alpha^d_{k \bm{p}'} .
   \nonumber\\ &&
\label{Hcollgb}
\end{eqnarray}
By making sure that all four gauge boson polarizations occur on an equal footing we have taken into account the Lorentz invariance of the free theory. The collisional contribution consists mostly of three-particle interactions, but the second-order term in $g$ represents a four-particle collision. Note that, for the construction of self-adjoint ghost particle operators, the Hamiltonian is self-adjoint with respect to the signed inner product.

\subsection{Evolution equations}
By recognizing $\iR \Qcommu{H}{A}$ as the time derivative of an operator $A$, we can now formulate the evolution equations for various operators and, in particular, we can compare them to the classical equations compiled in Section~\ref{secclassYM}. We begin with the evolution of the four-vector potential,
\begin{equation}\label{alphajevol}
    \iR \Qcommu{H}{\alpha^a_{j \bm{q}}} = - \varepsilon^a_{j \bm{q}}
    - \iR q_j \, \alpha^a_{0 \bm{q}}
    - g f^{abc} ( \alpha^b_j \alpha^c_0 )_{\bm{q}} ,
\end{equation}
which reassuringly is the quantum version of the Fourier transformed classical evolution equation (\ref{classYMevolA}). For the temporal component of the four-vector potential, we find
\begin{equation}\label{alpha0evol}
    \iR \Qcommu{H}{\alpha^a_{0 \bm{q}}} =
    \iR \sqrt{\frac{q}{2}} ( a^{0\, \dag}_{\bm{q}} + a^0_{-\bm{q}} ) =
    - \varepsilon^a_{0 \bm{q}}
    - \iR q_j \, \alpha^{a j}_{\bm{q}} ,
\end{equation}
which is the quantum version of the definition (\ref{E0convention}) of $\varepsilon^a_{0 \bm{q}}$, relating $\varepsilon^a_{0 \bm{q}}$ to the time derivative of $\alpha^a_{0 \bm{q}}$.

We next turn to the electric field type operators and find
\begin{eqnarray}
    \iR \Qcommu{H}{\varepsilon^a_{j \bm{q}}} &=& q^2 \alpha^a_{j \bm{q}}
    - q_j q_k \alpha^{a k}_{\bm{q}}
    + \iR q_j \varepsilon^a_{0 \bm{q}}
    + g f^{abc} \, (\alpha^b_0 \varepsilon^c_j)_{\bm{q}}
    \nonumber\\
    && \hspace{-5em} + \, \frac{\iR g f^{abc}}{\sqrt{V}} \!\!
    \sum_{\bm{q}',\bm{q}'' \in \bar{K}^3_\times}
    \delta_{\bm{q}'+\bm{q}'',\bm{q}}
    [( 2 q'_k + q''_k ) \, \alpha^b_{j \bm{q}'} - q'_j  \alpha^b_{k \bm{q}'} ] \alpha^c_{k \bm{q}''}
    \nonumber\\
    &-& \frac{g^2 f^{abs}f^{cds}}{V} \hspace{-1em}
    \sum_{\bm{p},\bm{p}',\bm{p}'' \in \bar{K}^3_\times}
    \delta_{\bm{p}+\bm{p}'+\bm{p}'',\bm{q}}
    \, \alpha^b_{k \bm{p}} \alpha^c_{k \bm{p}'} \alpha^d_{j \bm{p}''}    \nonumber\\
    &-& \frac{g f^{abc}}{\sqrt{V}} \sum_{\bm{q}',\bm{q}'' \in \bar{K}^3_\times}
    \delta_{\bm{q}'+\bm{q}'',\bm{q}} \, \iR q''_j \, \bar{c}^b_{\bm{q}'} c^c_{\bm{q}''} ,
\label{etajevol}
\end{eqnarray}
where the same simplification as in Eq.~(\ref{alpha0evol}) has been used. Except for the terms involving $\varepsilon^a_0$ and $\bar{c}^b c^c$, we recognize that exactly the same terms as in the classical evolution equation (\ref{classYMevolE}) in the absence of an external current. In view of Eq.~(\ref{E0convention}), $\varepsilon^a_0$ vanishes in the Lorenz gauge. Also the ghost term $\bar{c}^b c^c$ is clearly related to the proper handling of gauge conditions. The same remarks hold for the temporal component
\begin{equation}\label{eta0evol}
    \iR \Qcommu{H}{\varepsilon^a_{0 \bm{q}}} = \iR q_j \varepsilon^{a j}_{\bm{q}}
    + g f^{abc} \, ( \alpha^b_j \varepsilon^{c j} + \bar{c}^b \dot{c}^c )_{\bm{q}} ,
\end{equation}
which contains Eq.~(\ref{classYMevol1}) in the absence of an external current. The consistency of Eqs.~(\ref{alphajevol})--(\ref{eta0evol}) with the classical field equations implies that our minimal BRST invariant extension of the Lorentz covariant free theory indeed leads to the quantized Yang-Mills field theory.

For completeness, we also look at the evolution equations for the ghost operators,
\begin{equation}\label{cevol}
    \iR \Qcommu{H}{c^a_{\bm{q}}} = \dot{c}^a_{\bm{q}} ,
\end{equation}
and
\begin{equation}\label{cbarevol}
    \iR \Qcommu{H}{\bar{c}^a_{\bm{q}}} = \dot{\bar{c}}^a_{\bm{q}}
    + g f^{abc} \, (\alpha^b_0 \bar{c}^c)_{\bm{q}} .
\end{equation}
This last equation clearly shows that the operator $\dot{\bar{c}}^a_{\bm{q}}$ introduced in Eq.~(\ref{ghostfieldsdot}) is not the full time derivative of $\bar{c}^a_{\bm{q}}$, as pointed out before. For non-Abelian gauge theories, interactions between ghost particles and vector bosons occur. More of these interactions are implied by
\begin{eqnarray}
    \iR \Qcommu{H}{\dot{c}^a_{\bm{q}}} &=& - q^2 c^a_{\bm{q}}
    + g f^{abc} \, (\alpha^b_0 \dot{c}^c)_{\bm{q}} \nonumber\\
    &+& \frac{g f^{abc}}{\sqrt{V}} \sum_{\bm{q}',\bm{q}'' \in \bar{K}^3_\times}
    \delta_{\bm{q}'+\bm{q}'',\bm{q}} \, \iR q''_j \, \alpha^{b j}_{\bm{q}'} c^c_{\bm{q}''} ,
    \nonumber\\ &&
\label{cdotevol}
\end{eqnarray}
and
\begin{equation}\label{cbardotevol}
    \iR \Qcommu{H}{\dot{\bar{c}}^a_{\bm{q}}} = - q^2 \bar{c}^a_{\bm{q}}
    + \frac{g f^{abc}}{\sqrt{V}} \sum_{\bm{q}',\bm{q}'' \in \bar{K}^3_\times}
    \delta_{\bm{q}'+\bm{q}'',\bm{q}} \, \iR q_j \, \alpha^{b j}_{\bm{q}'} \bar{c}^c_{\bm{q}''} .
\end{equation}

\subsection{Additional conserved charge}\label{secaddconscharge}
It can easily be verified that an operator $R$ possessing the properties
\begin{equation}\label{Rtrial1}
   \iR \Qcommu{R}{c^a_{\bm{q}}} = c^a_{\bm{q}} , \qquad
   \iR \Qcommu{R}{\bar{c}^a_{\bm{q}}} = - \bar{c}^a_{\bm{q}} ,
\end{equation}
and
\begin{equation}\label{Rtrial2}
   \iR \Qcommu{R}{\dot{c}^a_{\bm{q}}} = \dot{c}^a_{\bm{q}} , \qquad
   \iR \Qcommu{R}{\dot{\bar{c}}^a_{\bm{q}}} = - \dot{\bar{c}}^a_{\bm{q}} ,
\end{equation}
as well as vanishing commutators with all gauge boson operators, commutes with the full Hamiltonian $H$. Such an operator $R$ can be interpreted as the generator of the symmetry associated with the rescaling of ghost operators with and without bars by reciprocal factors.

As suggested in Eq.~(2.17) of \cite{KugoOjima78b}, we have the following natural candidate for $R$,
\begin{equation}\label{Rtrial}
   R = \sum_{\bm{q} \in \bar{K}^3_\times}
   \big( c^a_{\bm{q}} \dot{\bar{c}}^a_{-\bm{q}}
   - \dot{c}^a_{\bm{q}} \bar{c}^a_{-\bm{q}} \big) .
\end{equation}
One can easily verify by means of the anticommutation relations (\ref{ccbardotanticomrel}) that this operator $R$ indeed leads to the properties in Eqs.~(\ref{Rtrial1}), (\ref{Rtrial2}). The simple form of $R$ (compared to Eq.~(2.17) of \cite{KugoOjima78b}) is a consequence of the fact that the dots don't indicate full time derivatives but rather the operators defined in Eq.~(\ref{ghostfieldsdot}) or Eqs.~(\ref{ghostfielddot1}) and (\ref{ghostfielddot2}).

We further find the commutator
\begin{equation}\label{RQcom}
   \iR \Qcommu{R}{Q} = Q .
\end{equation}
On the kernel of $Q$, the operator $R$ commutes also with the BRST charge. We can hence restrict the physical states to an eigenspace of $R$ (most conveniently with eigenvalue zero).

\section{Construction of physical states}
In Section~\ref{secFockspace}, we had introduced two inner products, the canonical and the signed ones, where the latter implies negative-norm states and hence is not truly an inner product. As the signed inner product serves for defining the physical bra- in terms of ket-vectors, we need to identify a subspace of the full Fock space on which the signed product becomes a true inner product.

If the ghost operators are introduced according to Eqs.~(\ref{ghostfield1})--(\ref{ghostfielddot2}), the BRST charge (\ref{BRSTcharge}) is self-adjoint with respect to the signed inner product, $Q^\ddag = Q$. However, the BRST charge is not self-adjoint with respect to the canonical inner product, $Q^\dag \neq Q$. We here decompose the Fock space into three mutually orthogonal spaces (in terms of the canonical inner product) and discuss the physical implications (associated with the signed inner product).

\subsection{Decomposition of states}
A decomposition of Fock space into three mutually orthogonal subspaces is achieved in terms of the images and kernels of the BRST operator $Q$ and of the co-BRST operator $Q^\dag$. Our discussion is based on Section~2.5 of \cite{vanHolten05ip}. A toy illustration of the essential features of the subsequent development is offered in Appendix~\ref{apptoyQ}.

\begin{figure}
\centerline{\includegraphics[width=7 cm]{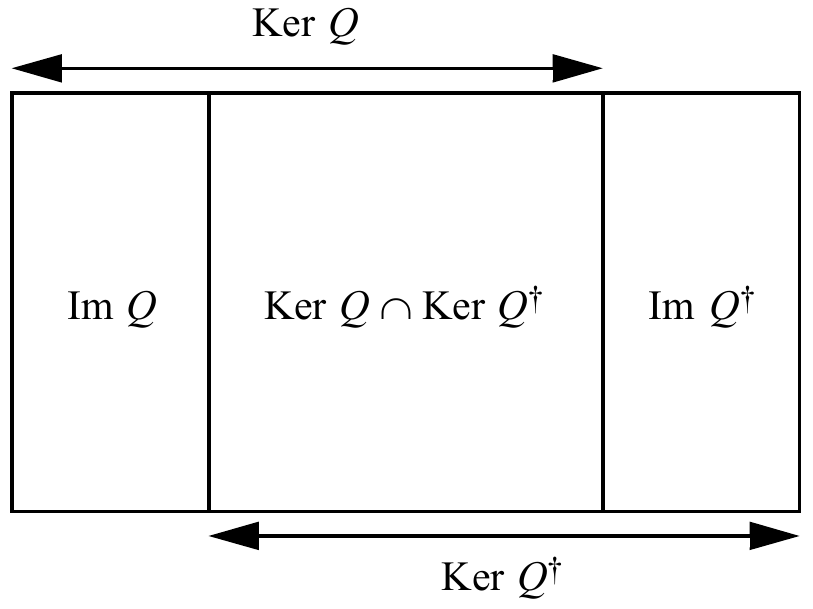}}
\caption[ ]{Decomposition of Fock space into three mutually orthogonal subspaces.} \label{BRST_YM_fig_decomp}
\end{figure}

Relying on the canonical inner product, the full Fock space ${\cal F}$ can be expressed as the direct sum of three mutually orthogonal spaces (the characteristic features of this decomposition are illustrated in Figure~\ref{BRST_YM_fig_decomp}),
\begin{equation}\label{Fockdecomposition}
   {\cal F} = ( {\rm Ker}\,Q \cap {\rm Ker}\,Q^\dag) \oplus {\rm Im}\,Q \oplus {\rm Im}\,Q^\dag .
\end{equation}
In view of the nilpotency properties $Q^2 = (Q^\dag)^2 =0$, we have
\begin{equation}\label{ImKerincl}
   {\rm Im}\,Q \subset {\rm Ker}\,Q , \qquad {\rm Im}\,Q^\dag \subset {\rm Ker}\,Q^\dag .
\end{equation}
As a next step, we prove the representations
\begin{equation}\label{Kerexpl}
   {\rm Ker}\,Q = ({\rm Im}\,Q^\dag)^\bot , \qquad {\rm Ker}\,Q^\dag = ({\rm Im}\,Q)^\bot .
\end{equation}
We only prove the first identity in Eq.~(\ref{Kerexpl}) because the second one can be shown in exactly the same way. If $\Dket{\varphi} \in {\rm Ker}\,Q$, that is $Q \Dket{\varphi} =0$, we have $0 = s^{\rm can}(Q \Dket{\varphi}, \Dket{\psi}) = s^{\rm can}(\Dket{\varphi}, Q^\dag \Dket{\psi})$ for all $\Dket{\psi} \in {\cal F}$. Conversely, if $\Dket{\varphi} \in {\cal F}$ is such that $s^{\rm can}(\Dket{\varphi}, Q^\dag \Dket{\psi})=0$ for all $\Dket{\psi} \in {\cal F}$, then $s^{\rm can}(Q \Dket{\varphi}, \Dket{\psi})=0$ for all $\Dket{\psi} \in {\cal F}$, which implies $Q \Dket{\varphi} = 0$.

Equations (\ref{ImKerincl}) and (\ref{Kerexpl}) imply that ${\rm Im}\,Q$ and ${\rm Im}\,Q^\dag$ are orthogonal spaces and that ${\rm Ker}\,Q \cap {\rm Ker}\,Q^\dag$ is orthogonal to both images and exhausts the rest of ${\cal F}$. We have thus established Eq.~(\ref{Fockdecomposition}) and the situation depicted in Figure~\ref{BRST_YM_fig_decomp}. This equation implies that every state $\Dket{\varphi} \in {\cal F}$ can be written as
\begin{equation}\label{vecdecompose}
   \Dket{\varphi} = \Dket{\chi} + Q \Dket{\psi_1} + Q^\dag \Dket{\psi_2} \quad {\rm with\ }
   \; Q \Dket{\chi} = Q^\dag \Dket{\chi} = 0 ,
\end{equation}
where the three contributions to $\Dket{\varphi}$ are mutually orthogonal in the canonical inner product.

\subsection{BRST Laplacian}
Before we discuss the signed norm of states in the various subspaces, we elaborate some details of the decomposition (\ref{Fockdecomposition}). This can be done in terms of the BRST Laplacian
\begin{equation}\label{BRSTLapldef}
   \Delta = (Q+Q^\dag)^2 = Q Q^\dag + Q^\dag Q .
\end{equation}
As $\Delta$ is the square of a self-adjoint operator, its eigenvalues $\lambda^2$ must be real and nonnegative. All states in ${\rm Ker}\,Q \cap {\rm Ker}\,Q^\dag$ are eigenstates of $\Delta$ with eigenvalue zero. A nonzero eigenvalue can only be obtained for vectors from one of the images, say $\Dket{\psi} \in {\rm Im}\,Q$. To obtain $\Delta \Dket{\psi} \neq 0$, $\Dket{\varphi} = Q^\dag \Dket{\psi}$ has to lie in $({\rm Ker}\,Q)^\bot = {\rm Im}\,Q^\dag$. Therefore, eigenstates of $\Delta$ with nonzero eigenvalues can only be obtained by flipping back and forth between the two images when applying the two factors in the definition of $\Delta$. By rescaling $\Dket{\varphi}$, we find the following doublet of states,
\begin{eqnarray}
   Q \Dket{\varphi} &=& \lambda \Dket{\psi} \nonumber\\
   Q^\dag \Dket{\psi} &=& \lambda \Dket{\varphi} .
\label{doublet}
\end{eqnarray}
For $\lambda=0$ we would end up in the kernel of $\Delta$, so that
\begin{equation}\label{kerDelta}
   {\rm Ker}\,\Delta = {\rm Ker}\,Q \cap {\rm Ker}\,Q^\dag .
\end{equation}

According to Eq.~(\ref{doublet}), both $\Dket{\varphi} \in {\rm Im}\,Q^\dag$ and $\Dket{\psi} \in {\rm Im}\,Q$ are eigenstates of $\Delta$ with the same eigenvalue $\lambda^2$. We further note the property
\begin{equation}\label{mixedImstates}
   (Q+Q^\dag) (\Dket{\varphi} \pm \Dket{\psi}) =
   \lambda (\Dket{\psi} \pm \Dket{\varphi}) ,
\end{equation}
so that $\Dket{\varphi} \pm \Dket{\psi}$ are eigenvectors of $Q+Q^\dag$ with eigenvalues $\pm \lambda$. In summary, we have shown that the eigenvectors of $\Delta$ with nonzero eigenvalues are nicely organized in doublets.

\subsection{Physical subspace}
As a next step, we wish to identify ${\rm Ker}\,\Delta = {\rm Ker}\,Q \cap {\rm Ker}\,Q^\dag$ as the physical subspace of ${\cal F}$ in which $s^{\rm sign}$ becomes a valid inner product, to be taken as the physical inner product. In other words, physical states are characterized by $Q \Dket{\varphi} = Q^\dag \Dket{\varphi} = 0$.

In the natural base of our Fock space, the canonical inner product is represented by the unit matrix, whereas the signed inner product is represented by a diagonal matrix $\sigma$ with diagonal elements $\pm 1$. If $A^\dag$ and $A^\ddag$ are the matrices representing the two adjoints of an operator $A$ in the natural basis, the definition of these adjoints implies $\sigma A^\dag = A^\ddag \sigma$ and, in view of $\sigma^2=1$, also $\sigma A^\ddag = A^\dag \sigma$. In view of the self-adjointness property $Q^\ddag=Q$, we conclude that we can choose a canonically orthonormal basis of eigenvectors of $Q+Q^\dag$ which all possess signed norm $+1$ or $-1$.

We note that the signed norm of any state from one of the images of $Q$ or $Q^\dag$ vanishes. For example, $s^{\rm sign}(Q \Dket{\varphi}, Q \Dket{\varphi}) = s^{\rm sign}(\Dket{\varphi}, Q^\ddag Q \Dket{\varphi}) = 0$ as $Q^\ddag = Q$ and $Q^2 = 0$. By superposition of states from the two images one can produce states of negative norm. For the eigenvectors of $Q+Q^\dag$ found in Eq.~(\ref{mixedImstates}), we have
\begin{equation}\label{posnegnormstates}
   s^{\rm sign}(\Dket{\varphi} \pm \Dket{\psi}, \Dket{\varphi} \pm \Dket{\psi}) =
   \pm ( \Dbraket{\varphi}{\psi} + \Dbraket{\psi}{\varphi}) .
\end{equation}
This result nicely shows that the (properly scaled) eigenvectors of $Q+Q^\dag$ come in pairs with signed norms $+1$ and $-1$.

Can we now guarantee that the signed inner product is positive on ${\rm Ker}\,\Delta = ({\rm Im}\,Q \oplus {\rm Im}\,Q^\dag)^\bot$? The toy example of Appendix~\ref{apptoyQ} shows that the answer is `no.' In the BRST construction we have to make sure that the number of negative-norm states matches exactly the pairs in ${\rm Im}\,Q \oplus {\rm Im}\,Q^\dag$. We need to verify that every negative-norm state can be written as a linear combination of zero-norm states from ${\rm Im}\,Q$ and ${\rm Im}\,Q^\dag$.

The above construction is usually presented in two steps. For the physical states, one first imposes the condition $Q \Dket{\varphi}=0$. By excluding ${\rm Im}\,Q^\dag$ one avoids the above construction of negative-norm states from the doublets (\ref{doublet}), but zero-norm states clearly still exist in ${\rm Im}\,Q$. In a second step, one considers equivalence classes of states that differ only by zero-norm states. Selecting representatives of the equivalence classes is often referred to as gauge fixing. Our gauge fixing condition thus is $Q^\dag \Dket{\varphi} = 0$. According to Eq.~(\ref{vecdecompose}), the states satisfying the first condition $Q \Dket{\varphi}=0$ possess the representation
\begin{equation}\label{vecdecomposes}
   \Dket{\varphi} = \Dket{\chi} + Q \Dket{\psi} \quad {\rm with\ }
   \; Q \Dket{\chi} = Q^\dag \Dket{\chi} = 0 .
\end{equation}
We can take $\Dket{\chi}$ as the unique representative of an equivalence class of states that differ by the zero-norm states $Q \Dket{\psi}$ for some $\Dket{\psi} \in {\cal F}$, so that the physical subspace indeed is ${\rm Ker}\,\Delta$. The physical norms are independent of the choice of the representative. For Hamiltonians commuting with $Q$ and $Q^\dag$, the physical subspace ${\rm Ker}\,\Delta$ is invariant under Hamiltonian dynamics.

\section{Inclusion of matter}
In the terms proportional to $g$ in the BRST charge (\ref{BRSTcharge}) we recognize a term that contains the temporal component (\ref{classfieldcurr0}) of the current four-vector resulting from the gauge bosons. The simplest way to incorporate matter into the BRST charge is to add the corresponding term involving the temporal component of the current four-vector associated with matter,
\begin{equation}\label{QJdef}
   Q_J = - \sum_{\bm{q} \in \bar{K}^3_\times} J^{a 0}_{\bm{q}} c^a_{-\bm{q}} .
\end{equation}
This idea is consistent with the expression for the BRST charge in quantum electrodynamics (see, for example, Eq.~(3.125) of \cite{hcoqft}). Nilpotency of the extended BRST charge $Q+Q_J$ requires
\begin{equation}\label{nilpotentmatter}
   Q_J^2 + \Qantico{Q}{Q_J} = 0 .
\end{equation}
Both terms can be calculated under the assumption that $J^{a 0}_{\bm{q}}$ commutes with all gauge boson and ghost operators:
\begin{equation}\label{nilpotentmatter1}
   \Qantico{Q}{Q_J} = \frac{\iR}{2} g f^{abc} \sum_{\bm{q} \in \bar{K}^3_\times}
   J^{c 0}_{\bm{q}} ( c^a c^b )_{-\bm{q}} ,
\end{equation}
and
\begin{equation}\label{nilpotentmatter2}
   Q_J^2 = \frac{1}{2} \sum_{\bm{q},\bm{q}' \in \bar{K}^3_\times}
   \Qcommu{J^{a 0}_{\bm{q}}}{J^{b 0}_{\bm{q}'}} \,
   c^a_{-\bm{q}} c^b_{-\bm{q}'} .
\end{equation}
After using Eq.~(\ref{convolution}), the nilpotency condition can be written as
\begin{equation}\label{currentalgebra0}
   \Qcommu{J^{a 0}_{\bm{q}}}{J^{b 0}_{\bm{q}'}} =
   - \iR \, \frac{g f^{abc}}{\sqrt{V}} \, J^{c 0}_{\bm{q} + \bm{q}'} .
\end{equation}

With this commutator we obtain the BRST transformation
\begin{equation}\label{BRSTtransformJ0}
   \delta J^{a 0}_{\bm{q}} = - g f^{abc} \, (J^{b 0} c^c)_{\bm{q}} ,
\end{equation}
which is exactly what we expect in view of the classical gauge transformation (\ref{gaugetransJ}). To recover also the proper transformation behavior of the spatial components of the current four-vector, we need to generalize Eq.~(\ref{currentalgebra0}) to
\begin{equation}\label{currentalgebramu}
   \Qcommu{J^{a 0}_{\bm{q}}}{J^{b \mu}_{\bm{q}'}} =
   - \iR \, \frac{g f^{abc}}{\sqrt{V}} \, J^{c \mu}_{\bm{q} + \bm{q}'} .
\end{equation}

Equation (\ref{currentalgebramu}) is a simple case of a current algebra, in which neither an axial current nor a Schwinger term is considered. Additional terms would require a change of the BRST charge and/or Hamiltonian to make sure that the BRST approach can still be used to handle the constraints associated with gauge theories. For example, the Schwinger term \cite{Schwinger59} has been discussed in the context of Yang-Mills theory in Eq.~(3.16) of \cite{deAzcarragaetal90} or Eq.~(1.6) of \cite{MickelssonRajeev88}. It is typically associated with sums that are not absolutely convergent so that regularization is required. For electromagnetic fields and massless fermions in one space dimension (known as the Schwinger model \cite{Schwinger62}), the Schwinger term is finite and well-defined. Possible modifications of the Hamiltonian and BRST charge are discussed in Section~3.3 of \cite{hcoqft}. As at least temporal photons have to acquire mass, the Schwinger term leads to chiral symmetry breaking.

If we choose the Hamiltonian for the interaction of gauge bosons and matter as
\begin{equation}\label{HamiltonianJ}
   H_J = - \sum_{\bm{q} \in \bar{K}^3_\times}
   J^{a \mu}_{-\bm{q}} \, \alpha^a_{\mu \bm{q}} ,
\end{equation}
we find
\begin{equation}\label{etamuevolm}
    \iR \Qcommu{H_J}{\varepsilon^a_{\mu \bm{q}}} = - J^a_{\mu \bm{q}} ,
\end{equation}
and hence the proper occurrence of the current in Eqs.~(\ref{classYMevol1}) and (\ref{classYMevol2}). In order to check whether $H+H_J$ is BRST invariant with respect to the new BRST charge $Q+Q_J$, we calculate the commutators
\begin{eqnarray}
    \iR\Qcommu{Q}{H_J} &=& \sum_{\bm{q} \in \bar{K}^3_\times}
    \big( J^a_{0 \bm{q}} \, \dot{c}^a_{-\bm{q}} -
    \iR q_j \, J^a_{j \bm{q}} \, c^a_{-\bm{q}} \big) \nonumber\\
    &-& g f^{abc} \sum_{\bm{q} \in \bar{K}^3_\times}
    (J^{b \mu} c^c)_{-\bm{q}} \, \alpha^a_{\mu \bm{q}} ,
\label{HJinBRST1}
\end{eqnarray}
\begin{equation}\label{HJinBRST2}
    \iR\Qcommu{Q_J}{H} = - \sum_{\bm{q} \in \bar{K}^3_\times}
     J^a_{0 \bm{q}} \, \dot{c}^a_{-\bm{q}} ,
\end{equation}
and
\begin{equation}\label{HJinBRST3}
    \iR\Qcommu{Q_J}{H_J} = g f^{abc} \sum_{\bm{q} \in \bar{K}^3_\times}
    (J^{b \mu} c^c)_{-\bm{q}} \, \alpha^a_{\mu \bm{q}} .
\end{equation}
The incomplete compensation of terms in Eqs.~(\ref{HJinBRST1})--(\ref{HJinBRST3}) implies that $H+H_J$ is not BRST invariant with respect to the new BRST charge $Q+Q_J$. This is not surprising as the Hamiltonian for non-interacting matter is still missing. BRST invariance of the full Hamiltonian requires
\begin{equation}\label{fullBRSTinv1}
    \iR\Qcommu{Q_J}{H_{\rm mat}^{\rm free}} = \sum_{\bm{q} \in \bar{K}^3_\times}
    \iR q_j \, J^a_{j \bm{q}} \, c^a_{-\bm{q}} ,
\end{equation}
or, by means of Eq.~(\ref{QJdef}),
\begin{equation}\label{fullBRSTinv2}
    \sum_{\bm{q} \in \bar{K}^3_\times}
    \Big( \iR\Qcommu{H_{\rm mat}^{\rm free}}{J^{a 0}_{\bm{q}}}
    - \iR q_j \, J^{a j}_{\bm{q}} \Big) \, c^a_{-\bm{q}} = 0 .
\end{equation}
The current four-vector must be defined such that the operator identity
\begin{equation}\label{fullBRSTinv3}
    \iR\Qcommu{H_{\rm mat}^{\rm free}}{J^{a 0}_{\bm{q}}}
    - \iR q_j \, J^{a j}_{\bm{q}} = 0
\end{equation}
is satisfied (see, e.g., Eq.~(3.95) of \cite{hcoqft} for quantum electrodynamics). In other words, the current four-vector must be constructed such that a continuity equation holds. However, the condition (\ref{fullBRSTinv3}) is not a complete balance equation, which would require occurrence of the full Hamiltonian instead of $H_{\rm mat}^{\rm free}$. An additional contribution
\begin{equation}\label{fullBRSTinv4}
    \iR\Qcommu{H_J}{J^{a 0}_{\bm{q}}} = g f^{abc} (\alpha^b_\mu J^{c \mu})_{\bm{q}}
\end{equation}
appears in the complete balance equation. In view of the classical continuity equation (\ref{classconti}) one might assume that this source term is compensated by including the current of the field. However, if we define the quantum version of the current four-vector,
\begin{equation}\label{J0gaugebos}
   \tilde{J}^{a 0}_{\bm{q}} = g f^{abc} ( \alpha^b_j \varepsilon^{c j} )_{\bm{q}} ,
\end{equation}
and
\begin{eqnarray}
   \tilde{J}^{a j}_{\bm{q}} &=& - g f^{abc} \Big[ (\alpha^b_0 \varepsilon^{c j})_{\bm{q}}
   + g f^{cde} ( \alpha^b_k \alpha^d_j \alpha^e_k )_{\bm{q}} \Big] \nonumber\\
   &+& \frac{\iR g f^{abc}}{\sqrt{V}}
   \sum_{\bm{q}',\bm{q}'' \in \bar{K}^3_\times}
   \delta_{\bm{q}'+\bm{q}'',\bm{q}} \,
   (q'_j  \alpha^b_{k \bm{q}'} - q'_k \alpha^b_{j \bm{q}'}) \, \alpha^c_{k \bm{q}''} , \nonumber\\
   &&
\label{Jjgaugebos}
\end{eqnarray}
a lengthy calculation shows that additional gauge terms appear,
\begin{eqnarray}
   \iR\Qcommu{H+H_J}{\tilde{J}^{a 0}_{\bm{q}}} - \iR q_j \, \tilde{J}^{a j}_{\bm{q}} &=&
   - g f^{abc} (\alpha^b_\mu J^{c \mu})_{\bm{q}} \nonumber\\
   && \hspace{-12em}+ \, \frac{g f^{abc}}{\sqrt{V}}
   \sum_{\bm{q}',\bm{q}'' \in \bar{K}^3_\times}
   \delta_{\bm{q}'+\bm{q}'',\bm{q}} \,
   \Big\{ \iR \Qcommu{H}{\varepsilon^b_{0 \bm{q}'}} \alpha^{c 0}_{\bm{q}''}
   - \iR q'_j \varepsilon^b_{0 \bm{q}'} \alpha^{c j}_{\bm{q}''} \nonumber\\
   && \hspace{-12em}- \, g f^{bde} \big[
   \dot{c}^d_{\bm{q}'} (\bar{c}^e \alpha^{c 0})_{\bm{q}''}
   - \iR q'_j c^d_{\bm{q}'} (\bar{c}^e \alpha^{c j})_{\bm{q}''} \big] \Big\} .
\label{continuity}
\end{eqnarray}
Therefore, a more careful analysis of say color conservation on the physical space is required.

\section{Summary and discussion}
We have elaborated all the details of the BRST quantization of Yang-Mills theory in a strictly Hamiltonian approach. A new representation of ghost-field operators in terms of canonical creation and annihilation operators has been introduced in Eqs.~(\ref{ghostfield1})--(\ref{ghostfielddot2}). This representation combines two pivotal advantages: (i) there exists a well-defined Fock space that serves as the underlying Hilbert space for carrying out the Hamiltonian approach and (ii) the BRST charge (\ref{BRSTcharge}) and the Hamiltonian (\ref{Htotal})--(\ref{Hcollgb}) turn out to be self-adjoint operators with respect to the physical inner product. The Fock space actually carries two inner products, a canonical and a signed one, where the restriction of the latter to a suitable subspace serves as the physical inner product.  The occurrence of negative-norm states in the unrestricted space explains why ghost particles can combine the anticommutation relations for fermions with zero spin, in an apparent violation of the spin-statistics theorem.

The construction of the BRST charge is based on the following two properties: (i) its role as the generator of the operator version of classical gauge transformations and (ii) its nilpotency. Lorentz symmetry is taken into account in the construction of the Fock space (with four gauge boson polarizations), in the formulation of the BRST charge (inherited from classical gauge transformations), and in the construction of the Hamiltonian as the minimal BRST invariant extension of the free theory (in which all four gauge boson polarizations are on an equal footing). The final Hamiltonian found in Eqs.~(\ref{Htotal})--(\ref{Hcollgb}) reproduces the field equations of classical Yang-Mills theory.

The interaction with matter has been included in terms of the current four-vector by adding the contributions (\ref{QJdef}) and (\ref{HamiltonianJ}) to the BRST charge and Hamiltonian, respectively. In order to make the BRST approach work, a current algebra has to be postulated, where a particularly simple one is given in Eq.~(\ref{currentalgebramu}). Any change in the current algebra, say by a Schwinger term, requires modifications of the BRST charge and the BRST invariant Hamiltonian. We have also discussed the formulation of the conservation law associated with BRST symmetry.

Whereas the Hamiltonian approach to BRST quantization on Fock space has the educational advantage of being nicely explicit and transparent, it has serious disadvantages in practical calculations. In particular, perturbation theory becomes very cumbersome (see Appendix~A of \cite{hcoqft} for some simplifications). Our motivation for elaborating the Hamiltonian approach stems from the formulation of dissipative quantum field theory \cite{hcoqft,hco200}, which is based on quantum master equations for evolving density matrices in time as an irreversible generalization of Hamiltonian dynamics. Dissipative quantum field theory, which is based on the idea that the elimination of degrees of freedom in renormalization procedures leads to the emergence of irreversibility, adds rigor, robustness and intuition to the field and hence has the potential to clarify the foundations of quantum field theory. Dissipation can easily be introduced in the Hamiltonian approach (see Section 1.2.3.2 of \cite{hcoqft}).

Stochastic simulation techniques developed for quantum master equations \cite{BreuerPetru} can then be used to simulate quantum field theories (see Sections 1.2.8.6 and 3.4.3.3 of \cite{hcoqft} for details). With the present paper, the simulation ideas so far applied only in a rudimentary way to quantum electrodynamics \cite{hco214}, become applicable to quantum chromodynamics.

\appendix

\section{Proof of $Q^2=0$}\label{appnilpotent}
In order to prove the nilpotency of the BRST charge defined in Eq.~(\ref{BRSTcharge}), we write $Q = Q_0 + Q_1 + Q_2 + Q_3$ with
\begin{equation}\label{BRSTchargeQ1}
   Q_0 = \sum_{\bm{q} \in \bar{K}^3_\times}
    \big( \varepsilon^a_{0 \bm{q}} \, \dot{c}^a_{-\bm{q}} -
    \iR q_j \, \varepsilon^{a j}_{\bm{q}} \, c^a_{-\bm{q}} \big) ,
\end{equation}
\begin{equation}\label{BRSTchargeQ2}
   Q_1 = g f^{abc} \sum_{\bm{q} \in \bar{K}^3_\times}
    ( \varepsilon^{a \mu} \alpha^b_\mu  )_{\bm{q}} \, c^c_{-\bm{q}} ,
\end{equation}
\begin{equation}\label{BRSTchargeQ3}
   Q_2 = - g f^{abc} \sum_{\bm{q} \in \bar{K}^3_\times}
    ( \dot{c}^a \bar{c}^b )_{\bm{q}} \, c^c_{-\bm{q}} ,
\end{equation}
and
\begin{equation}\label{BRSTchargeQ4}
   Q_3 = \frac{1}{2} g f^{abc} \sum_{\bm{q} \in \bar{K}^3_\times}
    \dot{\bar{c}}^a_{-\bm{q}} ( c^b c^c )_{\bm{q}} .
\end{equation}
Based on trivial canonical commutation and anticommutation relations, we find
\begin{equation}\label{nilpotentQ1Q1}
   Q_0^2 = 0 ,
\end{equation}
exactly as for Abelian gauge theory. The evaluation of $Q_1^2$ is based on the commutation relations (\ref{cancomrelaleta}) and (\ref{cancomrelaleta0}); the result is
\begin{equation}\label{nilpotentQ2Q2}
   Q_1^2 = - \iR g^2 f^{ads} f^{bcs} \sum_{\bm{q} \in \bar{K}^3_\times}
    ( \varepsilon^{d \mu} \alpha^c_\mu  )_{-\bm{q}} \, ( c^b c^a )_{\bm{q}} .
\end{equation}
By means of Eq.~(\ref{ccbardotanticomrel}) we find
\begin{equation}\label{nilpotentQ3Q3}
   Q_2^2 = \iR g^2 f^{cds} f^{abs} \sum_{\bm{q} \in \bar{K}^3_\times}
    ( \bar{c}^a \dot{c}^d )_{\bm{q}} ( c^b c^c )_{-\bm{q}} .
\end{equation}
To obtain
\begin{equation}\label{nilpotentQ4Q4}
   Q_3^2 = 0 ,
\end{equation}
we need the Jacobi identity (\ref{Jacobi}).

As a consequence of
\begin{equation}\label{nilpotentQ2Q3}
   \Qantico{Q_1}{Q_2} = 0 ,
\end{equation}
the nonzero contribution to $Q_1^2$ in  Eq.~(\ref{nilpotentQ2Q2}) can only be compensated by
\begin{equation}\label{nilpotentQ2Q4}
   \Qantico{Q_1}{Q_3} = -\frac{\iR}{2} g^2 f^{cds} f^{abs} \sum_{\bm{q} \in \bar{K}^3_\times}
    ( \varepsilon^{d \mu} \alpha^c_\mu )_{-\bm{q}} \, ( c^b c^a )_{\bm{q}} .
\end{equation}
Indeed, the Jacobi identity (\ref{Jacobi}) implies
\begin{equation}\label{nilpotentpartsum1}
   Q_1^2 + \Qantico{Q_1}{Q_3} = 0 .
\end{equation}
The nonzero contribution to $Q_2^2$ in  Eq.~(\ref{nilpotentQ3Q3}) can only be compensated by
\begin{equation}\label{nilpotentQ3Q4}
   \Qantico{Q_2}{Q_3} = \frac{\iR}{2} g^2 f^{ads} f^{bcs} \sum_{\bm{q} \in \bar{K}^3_\times}
    ( \bar{c}^a \dot{c}^d )_{\bm{q}} ( c^b c^c )_{-\bm{q}} .
\end{equation}
By once more using the Jacobi identity (\ref{Jacobi}), we find
\begin{equation}\label{nilpotentpartsum2}
   Q_2^2 + \Qantico{Q_2}{Q_3} = 0 .
\end{equation}

We still need to evaluate the anticommutators of $Q_0$ with all the other contributions to $Q$. We first evaluate
\begin{eqnarray}
   \Qantico{Q_0}{Q_1} &=& \iR g f^{abc} \sum_{\bm{q} \in \bar{K}^3_\times}
   \varepsilon^a_{0 \bm{q}} \, ( \dot{c}^b c^c )_{-\bm{q}} \nonumber\\
   &+& \frac{1}{2} g f^{abc} \sum_{\bm{q} \in \bar{K}^3_\times}
   q_j \varepsilon^{a j}_{\bm{q}} \, ( c^b c^c )_{-\bm{q}} . \qquad
\label{nilpotentQ1Q2}
\end{eqnarray}
The next anticommutator is given by
\begin{equation}\label{nilpotentQ1Q3}
   \Qantico{Q_0}{Q_2} = - \iR g f^{abc} \sum_{\bm{q} \in \bar{K}^3_\times}
   \varepsilon^a_{0 \bm{q}} \, ( \dot{c}^b c^c )_{-\bm{q}} ,
\end{equation}
which cancels the first term on the right-hand-side of Eq.~(\ref{nilpotentQ1Q2}). A final straightforward calculation yields
\begin{equation}\label{nilpotentQ1Q4}
   \Qantico{Q_0}{Q_3} = - \frac{1}{2} g f^{abc} \sum_{\bm{q} \in \bar{K}^3_\times}
   q_j \varepsilon^{a j}_{\bm{q}} \, ( c^b c^c )_{-\bm{q}} .
\end{equation}
As $\Qantico{Q_0}{Q_3}$ cancels the second term on the right-hand-side of Eq.~(\ref{nilpotentQ1Q2}), we obtain
\begin{equation}\label{nilpotentpartsum3}
   \Qantico{Q_0}{Q_1} + \Qantico{Q_0}{Q_2} + \Qantico{Q_0}{Q_3} = 0 .
\end{equation}

By summing up Eqs.~(\ref{nilpotentpartsum1}), (\ref{nilpotentpartsum2}), (\ref{nilpotentpartsum3}) and using the results (\ref{nilpotentQ1Q1}), (\ref{nilpotentQ4Q4}), (\ref{nilpotentQ2Q3}), we indeed obtain the desired nilpotency of the BRST charge defined in Eq.~(\ref{BRSTcharge}), $Q^2=0$.

\section{Construction of physical states -- a toy version}\label{apptoyQ}
We here sketch a toy version of the construction of physical states in the BRST approach. The general idea is nicely illustrated in a three-dimensional cartoon version of temporal, longitudinal and transverse gauge bosons (where physical states don't contain any right bosons, equivalent physical states differ by left bosons, and unique representatives of equivalence classes do not contain any left photons).

For the linear operators
\begin{equation}\label{QQdagtoy}
   Q = \left(
         \begin{array}{rrr}
           1 & -1 & 0 \\
           1 & -1 & 0 \\
           0 & 0 & 0 \\
         \end{array}
       \right) , \quad
   Q^\dag = \left(
         \begin{array}{rrr}
           1 & 1 & 0 \\
           -1 & -1 & 0 \\
           0 & 0 & 0 \\
         \end{array}
       \right) ,
\end{equation}
one easily verifies $Q^2 = (Q^\dag)^2 =0$. The image and kernel of the operator $Q$ are given by
\begin{equation}\label{ImQtoy}
   {\rm Im}\,Q = \left\{\left. \lambda \left( \begin{array}{r}
           1 \\
           1 \\
           0 \\
         \end{array} \right) \right| \lambda \in \mathbb{R} \right\} ,
\end{equation}
and
\begin{equation}\label{KerQtoy}
   {\rm Ker}\,Q = \left\{\left. \lambda \left( \begin{array}{r}
           1 \\
           1 \\
           0 \\
         \end{array} \right)
         + \lambda' \left( \begin{array}{r}
           0 \\
           0 \\
           1 \\
         \end{array} \right) \right| \lambda, \lambda' \in \mathbb{R} \right\} ,
\end{equation}
illustrating the general relation ${\rm Im}\,Q \subset {\rm Ker}\,Q$. We similarly have
\begin{equation}\label{ImQdagtoy}
   {\rm Im}\,Q^\dag = \left\{\left. \lambda \left( \begin{array}{r}
           1 \\
           -1 \\
           0 \\
         \end{array} \right) \right| \lambda \in \mathbb{R} \right\} ,
\end{equation}
\begin{equation}\label{KerQdagtoy}
   {\rm Ker}\,Q^\dag = \left\{\left. \lambda \left( \begin{array}{r}
           1 \\
           -1 \\
           0 \\
         \end{array} \right)
         + \lambda' \left( \begin{array}{r}
           0 \\
           0 \\
           1 \\
         \end{array} \right) \right| \lambda, \lambda' \in \mathbb{R} \right\} ,
\end{equation}
and ${\rm Im}\,Q^\dag \subset {\rm Ker}\,Q^\dag$. We also find ${\rm Im}\,Q^\dag = ({\rm Ker}\,Q)^\bot$ and ${\rm Im}\,Q = ({\rm Ker}\,Q^\dag)^\bot$, implying the orthogonality of ${\rm Im}\,Q$ and ${\rm Im}\,Q^\dag$. The kernel of $\Delta = (Q+Q^\dag)^2 = Q Q^\dag + Q^\dag Q$,
\begin{equation}\label{Laplaciantoy}
   \Delta = \left(
         \begin{array}{rrr}
           4 & 0 & 0 \\
           0 & 4 & 0 \\
           0 & 0 & 0 \\
         \end{array}
       \right) ,
\end{equation}
coincides with ${\rm Ker}\,Q \cap {\rm Ker}\,Q^\dag$, so that ${\rm Ker}\,\Delta$, ${\rm Im}\,Q$ and ${\rm Im}\,Q^\dag$ are three mutually orthogonal spaces. The total space is the direct sum of these three vector spaces. Any vector can uniquely be written as the sum of three vectors, one from each of these spaces.

Introducing a signed inner product by
\begin{equation}\label{sigmatoy}
   \sigma = \left(
         \begin{array}{rrr}
           -1 & 0 & 0 \\
           0 & 1 & 0 \\
           0 & 0 & 1 \\
         \end{array}
       \right) ,
\end{equation}
we realize the self-adjointness property $Q^\ddag = \sigma Q^\dag \sigma = Q$. For this signed inner product, states from ${\rm Im}\,Q$ or ${\rm Im}\,Q^\dag$ have zero norm, states from ${\rm Ker}\,\Delta$ have positive norm. Had we chosen a signed inner product with
\begin{equation}\label{sigmatoyalt}
   \sigma' = \left(
         \begin{array}{rrr}
           -1 & 0 & 0 \\
           0 & 1 & 0 \\
           0 & 0 & -1 \\
         \end{array}
       \right) ,
\end{equation}
we would still have the self-adjointness property $Q^\ddag = \sigma' Q^\dag \sigma' = Q$. However, we would have introduced too many states with a negative norm leading to an unphysical inner product on ${\rm Ker}\,\Delta$.

%\bibliography{hcopubs}

\end{document}